\documentclass[conference,12pt,onecolumn,draftclsnofoot]{IEEEtran}
\pdfoutput=1
\usepackage{epstopdf}
\usepackage{bm}
\usepackage{algorithm,algorithmic}
\usepackage{amsmath}
\usepackage{amssymb}
\usepackage{graphicx}
\ifCLASSINFOpdf
\else
\fi

\begin{document}
%
\title{A Novel Move-To-Front-or-Logarithmic Position (MFLP) Online List Update Algorithm }

\author{\IEEEauthorblockN{Baisakh}
\IEEEauthorblockA{Department of Computer Science\\ and Engineering\\
Veer Surendra Sai University of Technology\\
Burla, Sambalpur, India 768019\\
}
\and
\IEEEauthorblockN{Rakesh Mohanty}
\IEEEauthorblockA{Department of Computer Science\\ and Engineering\\
Veer Surendra Sai University of Technology\\
Burla, Sambalpur, India 768019}
}
\maketitle

\begin{abstract}
In this paper we propose a novel online deterministic list update algorithm known as \emph{Move-To-Front-or-Logarithmic Position (MFLP)}. Our proposed algorithm \emph{MFLP} achieves a competitive ratio of 2 for larger list with respect to static optimum offline algorithm, whereas \emph{MFLP} is not competitive for smaller list. We also show that \emph{MFLP} is \emph{2-competitive} with respect to dynamic optimum offline algorithm. Our results open up a new direction of work towards classifying the online algorithms based on competitive and not competitive.

\end{abstract}


%
\IEEEpeerreviewmaketitle

\section{Introduction} 
\subsection{List Update Problem} 
The \emph{List Update Problem (LUP)} proposed by \emph{McCabe} \cite{McCabe} is considered to be one of the classical online problem. The problem has been  extensively studied in the literature \cite{Bentley,Rivest,Binter,Gonnet,Binter1} since last five decades. The problem consists of two different inputs i.e an unsorted linear linked list \emph{L} and a sequence of requests $\sigma$. The list \emph{L} contains a set of distinct \emph{l} items i.e $<x_{1} , x_{2} , ..., x_{l}>$, where $x_{i}$ $\neq$ $x_{i+1}$ and $i \in {1, 2, .., l}$. These set of items need to be maintained by the list update algorithm while serving a sequence of requests $\sigma=$ $ <\sigma_{1}, \sigma_{2}, ..., \sigma_{n} >$. A single request $\sigma_{i}$ refers to an access operation to the requested item present in the list where  $\sigma_{i} \in L$. Every time an item is accessed, the list configuration can be changed by performing the list reorganization through exchange and reordering of items of  the list. The standard cost model proposed by \emph{Sleator and Tarjon} \cite{Sleator} defines the access and the list reorganization cost.

In the standard cost model, accessing an item $x_{i}$ at the $i^{th}$ position of the list incurs a cost of \emph{i}. Once an item is accessed it can be moved to any position forward of the list with no extra cost which is known as a free exchange. But if any two consecutive items are exchanged with respect to their relative positions, then it incurs a cost of \emph{1}, and is known as a paid exchange. The objective of the list update algorithm is to minimize the sum of access cost and reorganization cost while serving a request sequence $\sigma$.

\subsection{Online Algorithms} In an online scenario, a list update algorithm serves every request without having the knowledge of future requests. Such an algorithm is known as the online list update algorithm and is denoted by $ALG$. At every time instance $t_{i}$ and input $\sigma_{i}$ is known to the algorithm where as the rest of requested items i.e $< \sigma_{i+1}, \sigma_{i+2} , ..., \sigma_{n}>$ are unknown. 

Though many deterministic online list update algorithms have been studied in the literature \cite{Bachrach}, most of the algorithms are the variants of three well known primitive algorithms. They are \emph{Move-To-Front (MTF)}, \emph{Transpostion (TRANS)} and \emph{Frequency Count (FC)} as follows.\\
\emph{\bf{Move-To-Front (MTF):}} An item recently been accessed, is moved to the front of the list.\\  
\emph{\bf{Transpose (TRANS):}} An immediately accessed item is moved to one position forward in the list by transposing with its immediately preceding item.\\  
\emph{ \bf{Frequency Count (FC):}} When an item is accessed, the count value is incremented to one and then sort the items in the list according to the decreasing order of their frequency.\\  
 Among these three algorithms, \emph{MTF} is most widely used in practical applications and well studied in the literature.

\subsection{Competitive Analysis and Competitive Ratio} 
Competitive analysis \cite{Manasse} is one of the standard performance measures for the online algorithms \cite{Borodin,Borodin1, Fiat} . Competitive analysis is performed by comparing the worst case cost incurred by an online algorithm \emph{ALG} over all finite request sequences, with the cost incurred by the optimum offline algorithm \emph{OPT}. Optimum offline algorithm gives the minimum cost while serving any given request sequence $\sigma$. Let the cost incurred by $ALG$ and $OPT$ be denoted by $C_{ALG}(L,\sigma)$ and $C_{OPT}(L,\sigma)$ respectively. Then an online algorithm \emph{ALG} is called \emph{d-competitive} if there exist a constant $\beta$ such that $C_{ALG}(L,\sigma)$ $\leq$ $d.C_{OPT}(L,\sigma)$ $+\beta$ for all request sequences of $\sigma$. The less the competitive ratio the better the performance of an online algorithm.

\subsection{Practical and Research Motivation} 
Data compression is one of the most common and widely used application of the \emph{LUP}\cite{Albers} \cite{Dorrigiv}. \emph{LUP} also finds applications in maintaining dictionary, symbol table for compiler and finding point maxima in convex hull. Offline list update problem has been proved to be the $NP\-hard$ in \cite{Ambuhl}. \emph{Move-To-Front (MTF)} has been shown to be the best known online list update algorithm in the literature for practical applications \cite{Angelopoulos}, and has obtained the best competitive bound. It is a major research challenge to design competitive online list update algorithms which can outperform \emph{MTF} for real life inputs.


%

\section{Study of Related Work and Their Analysis}
\subsection{Known Previous Results:} Design of a better competitive online algorithm is achievable through the better understanding of optimum offline algorithm. Often we consider the static optimum offline \emph{STAT} algorithm while performing the competitive analysis of an online algorithm. In \emph{STAT}, items are arranged according to their non increasing order of their frequency count at the beginning and then the list remains unchanged till all items are served.

There are three primitive online deterministic list update algorithms which are widely used in the literature, such as \emph{Move-To-Front (MFM)}, \emph{Frequency Count (FC)} and \emph{Transpose (TRANS)}. Design of variants of these three standard algorithms and analysing their performance through competitive analysis have been the central interests of the researchers. All these studies have shown that some of the online algorithms are competitive and some are not competitive with respect to \emph{STAT}. Bentley and McGeoch \cite{Bentley} shown that \emph{Move-To-Front (MTF)} is \emph{2-competitive} with respect to static optimum offline algorithm \emph{(STAT)}. Whereas \emph{TRANS} is not competitive. In our recent work, we have shown that \emph{Move-To-Front-or-Middle (MFM)} is \emph{2-competitive} with respect to dynamic optimum offline algorithm, but not \emph{2-competitive} with respect to \emph{STAT}. In this paper we attempt to find a bound that separates \emph{competitive} and \emph{non-competitive} online algorithms with respect to \emph{STAT}.  

\subsection{Our Analysis of  Existing Algorithms:}
In this section, we have made a few observations by doing the analysis of existing \emph{MTF} and \emph{MFM} \cite{Mohanty} online algorithms. The following analysis not only depict about the competitiveness of online algorithm but also seek underlying reasons that make one algorithm as competitive and other as not competitive with rspect to \emph{STAT}.

\emph{\bf Analysis of MTF:} Lets consider a given list \emph{L} of size \emph{l} that contains the items $<x_{1}, x_{2},..x_{m},\\ x_{m+1},.., x_{l-1},x_{l}>$. An item $x_{i}$ represents the $i^{th}$ item of the list \emph{L}, where $x_{i} \in \{1,2,.., m,..,l \}$ and \emph{m} denotes the middle position of the list \emph{L}. Let $\sigma$ be the cruel request sequence which contains all the items of the list \emph{L} such that $\sigma=$ $(<x_{l},x_{l-1},.., x_{m}, x_{m-1},..x_{2}, x_{1}>)^{k}$ of size \emph{n}, where each \emph{l} items are repeated \emph{k} times. Since all the items in $\sigma$ are present in the reverse order to that of the list configuration, every access of an item in $\sigma$ by \emph{MTF} incurs a cost \emph{l}. So the total access cost incurred by \emph{MTF} is $C_{MTF}(L,\sigma)$ is $k(l*l)$. The total access cost incurred by \emph{STAT} for the given $\sigma$ is $C_{STAT}(L,\sigma)$ is $k(\sum_{i=1}^{l}(i))$. The competitive ratio is $\frac{C_{MTF}(L,\sigma)}{C_{STAT}(L,\sigma)} = \frac{(l*l)}{\sum_{i=1}^{l}(i)} \leq 2$. 

\emph{\bf Analysis of MFM:} Let the initial configuration of the list \emph{L} be $<x_{1}, x_{2},.., x_{m},x_{m+1},..., x_{l}>$ and the cruel request sequence be $\sigma$ of size \emph{n}. Here we have considered the cruel request sequence for \emph{MFM} with the assumption that all items are involved in $\sigma$. Here a subsequence of request sequence dominates $\sigma$ by being repeated \emph{k} times. Let the cruel request sequence $\sigma$ is presented as $<x_{m}, x_{m-1},.., x_{2}, (x_{l}, x_{l-1},..,x_{m+2}, x_{m+1}, x_{1})^k >$, where there are $(l-m+1)$ items are repeated \emph{k} times in the subsequence of $\sigma$. For a large request sequence the access cost of $<x_{m}, x_{m-1},.., x_{2}>$ are negligible to the total access cost incurred by \emph{MFM} to serve $\sigma$. Hence we only consider the subsequence $<(x_{l}, x_{l-1},..,x_{m+2}, x_{m+1}, x_{1})^k >$ as the cruel request sequence which has been repeated \emph{k} times in $\sigma$. So the total access cost incurred by \emph{MFM} to serve $\sigma$ is $C_{MFM}(L,\sigma)$ is $k[l(l-m+1)]$.

Since the optimum offline algorithm $\emph{STAT}$ perform the paid exchanges to bring \emph{(l-m+1)} items to the front, it incurs some cost. But as compared to the total cost for a large value of \emph{n}, its very small. So we ignore the paid cost while considering the total access cost to serve the cruel request sequence $\sigma$. Hence the total access cost incurred by \emph{STAT} to serve $\sigma$ is $k(\sum_{i=1}^{l-m+1}(i))$. The competitive ratio is  $\frac{C_{MTF}(L,\sigma)}{C_{STAT}(L,\sigma)} = \frac{l(l-m+1)}{\sum_{i=1}^{l-m+1}(i)} \geq 2$. We use this competitive ratio in the \emph{Theorem-3} to show that \emph{MFM} converges to 4.

\emph{\bf Observation:} We observed that due to the nature of \emph{MTF}, all the items are involved in the cruel request sequence. Here both the numerator and denominator in the competitive ratio are associated with \emph{l} items which makes \emph{MTF} as \emph{2-competitive}. But in case of \emph{MFM}, $l-m+1$ items are involved in both numerator and denominator, which results \emph{MFM} to be not \emph{2-competitive}. So fixing up a bound on the involvement of elements to produce \emph{2-competitive} algorithm is challenging. Here we raise an open question that for what value of \emph{p}, an online algorithm is \emph{2-competitive} where the value of \emph{p} may vary from \emph{2 to m}?
  
\emph{\bf Motivation} In this paper, we make our first attempt to address this above question and we propose an online deterministic algorithm known as \emph{Move-To-Front-or-Logarithmic-Position (MFLP)}. We  show that when the value of \emph{p} is $log_{2}L$, where $p<<m$, it is \emph{2-competitive for larger size list} with respect to $STAT$. In this scenario the competitive ratio of \emph{MFLP} is converging to 2. But for smaller size it is not \emph{2-competitive}. Since the nature of static optimum offline algorithm is fixed, here we introduce a new parameter that is the number of elements involved in constructing the cruel request sequence $\sigma$ while performing competitive analysis. Our results open up a new direction of work to classify online algorithms based competitive and non competitive algorithms with respect to \emph{STAT} such as \emph{2-competitive, 3-competitive or 4-competitive} online algorithms.

\section{Our Proposed Move-To-Front-or-Logarithmic Position (MFLP)  Algorithm }
Our Proposed \emph{Move-To-Front-or-Logarithmic-Position (MFLP)} algorithm has been motivated by the \emph{Move-To-Front-Middle (MFM)} algorithm \cite{Mohanty} which was experimentally shown to be the better algorithm with respect to \emph{Move-To-Front (MFM)} on the \emph{Calgary Corpus} and \emph{Canterbury Corpus} data sets.  The idea behind this strategy is not to bring the accessed item to the front of the list if the accessed item is present behind the middle position of the list. This strategy suits well for the request sequences which are having less degree of locality of reference for a small size of list. If the degree of locality reference increases with the increase in size of the list \emph{L}, then performance of \emph{MFM} is not as good as the performance of \emph{MTF}. But interestingly our proposed algorithm \emph{MFLP} performs better than \emph{MFM} as the list size grows. In this paper, we have performed theoretical studied on our proposed algorithm \emph{MFLP} for a large size list and shown that the competitive ratio of  \emph{MFLP} is converging to 2. We have also shown that \emph{MFLP} is \emph{2-competitive} with respect to dynamic optimum offline algorithm denoted as \emph{OPT}.

\emph{\bf Move-To-Front-or Logarithmic-Position (MFLP):} When an item \emph{x} is accessed, move \emph{x} to the $p^{th}$ position of the list \emph{L}, where $p= \lceil log_{2}L \rceil$ if the position of the \emph{x} is greater then \emph{p}, else move \emph{x} to the front of the list.

\subsection{Basic terms and Definitions}

Let \emph{L} be the list of \emph{l} distinct items repressed as $<x_{1}, x_{2}, ..., x_{p},..., x_{l}>$ and $\sigma$ be the request sequence of \emph{n} items which can be represented as $\sigma =<\sigma_{1},\sigma_{2},....,\sigma_{n}>$. Let $\sigma_{i}$ represents the $i^{th}$ item in the request sequence $\sigma$ where $\sigma_{i} \in {x_{1}, x_{2}, ..., x_{n}}$ and $1 \leq i \leq n$.
$x_{p}$ be the logarithmic position of the list \emph{L}, where $p=$ $\lceil{log_{2} l}\rceil$. The total access cost incurred by \emph{MFLP} to serve $\sigma$ on the given list \emph{L} is $C_{MFLP}(L,\sigma)$.

 	
\subsection{Illustration of MFLP Algorithm}
The nature of \emph{Move-To-Front-or-Logarithmic-Position (MFLP)} online algorithm is illustrated in the \emph{figure-1}. We consider a list \emph{L} with the initial configuration $<1,2,3,4,5,6>$ and a request sequence $\sigma=$ $<2,4,1,5,3,6>$. Let \emph{p} be the logarithmic position of the list \emph{L} i.e $p=$ $\lceil{log_{2} L}\rceil$ $=$ $\lceil{log_{2} 6 }\rceil$ $=3$. When the items 2 and 1 are accessed, their respective positions of the list are less than \emph{p}, hence both the items are moved to the front of the list. Whereas for the remaining items present in $\sigma$, items are moved to the $p^{th}$ position of the list immediately after the items are accessed. The total cost incurred by \emph{MFLP} to $\sigma$ i.e $C_{MFM}(l,\sigma)$ is $24$. We observe that for this request sequence \emph{MTF} incurs the total cost i.e $C_{MTF}(l,\sigma)= 25$. Since the request sequence contains no locality of reference, \emph{MFLP} gives better cost than \emph{MTF}.

\begin{figure}
\centering
\includegraphics[width=8cm,height=7cm]{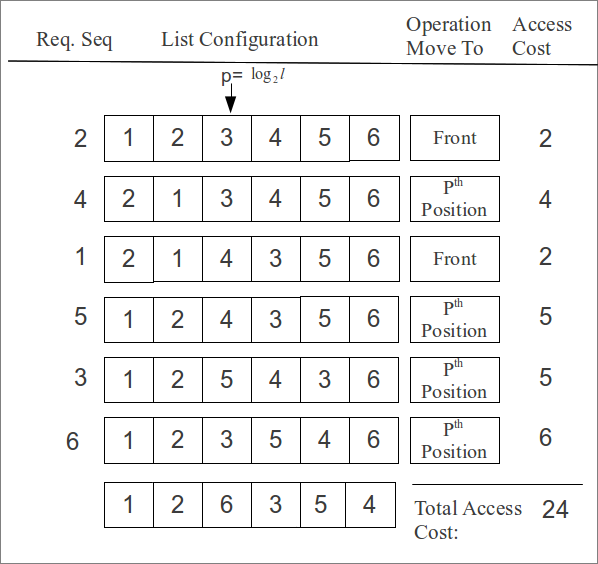} 
\caption{Illustration of MFLP Algorithm}
\label{fig:mflpc.PNG}
\end{figure}

\section{Competitive Analysis of Move-To-Front-or-Logarithmic-Position (MFLP) Algorithm}
In this section, we investigate the competitive bound for the $MFLP$ algorithm. We construct the cruel request sequence $\sigma$ for the \emph{MFLP} and observe the behaviour of \emph{MFLP} on different size of list \emph{L}. This observation has motivated us to design two different optimal offline strategies which could ensure to give an upper bound on the total optimal cost.
In our analysis we consider two different optimal offline strategies i.e the $\emph{static optimal offline algorithm}(STAT)$ and the $\emph{dynamic optimal offline algorithm (OPT)}$.  The static optimum offline algorithm is denoted as \emph{STAT} which initially reorganizes the list before serving the request sequence and no further reorganization is done while serving the request sequence $\sigma$. The reorganization is done on the basis of decreasing frequency of the requested items in $\sigma$ by using only the minimum paid exchanges. The dynamic optimum offline algorithm, known as \emph{OPT} performs dynamic list reorganization before or after serving any request in the request sequence using paid exchanges.\\

{\bf Theorem:1} \emph{\bf  \emph{MFLP} is not \emph{2-competitive} with respect to STAT}.

{\bf Theorem:2} \emph{\bf  The competitive ratio of \emph{MFLP} converges to 4 for larger size list with respect to STAT}.

{\bf Proof:} In the competitive analysis of \emph{MFLP}, we consider the same assumption, which we have  made for \emph{MFM} analysis in the previous section. Let the initial list \emph{L} be considered as $<x_{1}, x_{2},..x_{p},x_{p+1},..,x_{l-1},x_{l}>$, where the \emph{p} represents the logarithmic position of the list i.e $p=\lceil{log_{2} l}\rceil$. Let the cruel request sequence $\sigma$ for \emph{MFLP} be $<(x_{l}, x_{l-1},..,x_{p+2}, x_{p+1}, x_{p})^k >$ where $l-p+1$ items are repeated \emph{k} times. The cost incurred to serve $\sigma$  by \emph{STAT} is $C_{STAT}(L,\sigma)$ $\sum_{i=1}^{l-p+1}(i)$. To serve the given $\sigma$, each element would incur an access cost as \emph{l}. Hence the total access cost incurred by \emph{MFLP} to serve $\sigma$ is $C_{MFL}(L,\sigma)$ $l(l-p+1)$. 

Hence the competitive ratio is $\frac{C_{MFLP}(L,\sigma)}{C_{STAT}(L,\sigma)}=$ $\frac{k[l(l-p+1)]}{k[\sum_{i=1}^{l-p+1}(i)]}$. When the value of \emph{L} tends to infinity then the competitive ratio becomes i.e $\lim_{l\to\infty}({\frac{l(l-p+1)}{\sum_{i=1}^{l-p+1}(i)}})$ = $\lim_{l\to\infty}({\frac{l(l-log_{2}l+1)}{\frac{(l-log_{2}L+1)(l-Log_{2}l+2)}{2}}})$ $=$ $\lim_{l\to\infty} (\frac{2l}{l-log_{2}l+2})$.
By dividing the factor \emph{l} both in numerator and denominator, the competitive ratio simplifies to\\ $\lim_{l \to \infty} (\frac{2}{1- \frac{log_{2}l}{l} + \frac{2}{l}})$ $\approx$ $2$.\\

\emph{\bf Our Observations:} Here we analyse the behaviour of \emph{MFLP} for dynamic list by varying the list size from $2^{2}$ to $2^{50}$. Our analysis also involve the nature of deviation of \emph{MFLP} and \emph{MFM} algorithm from the best known \emph{MTF} algorithm which we have shown in terms of table and figure.
\begin{figure}[!ht]
\centering
  \includegraphics[width=13cm,height=10cm]{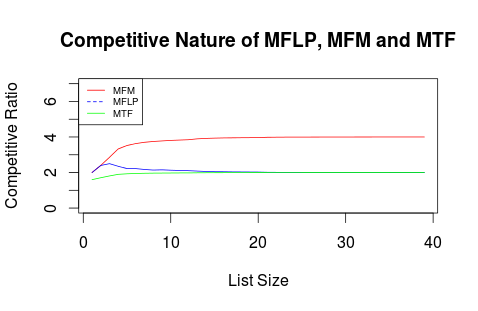}
  \caption{Competitive Nature of MFLP, MFM and MTF}
  \label{fig: mflp.PNG}
 \end{figure} 

From the \emph{figure-2}, it is observed that when the list size grows to 40, the competitive ratio of \emph{MFLP} starts falling down. And gradually as the list size grows, it converges with the line of \emph{MTF} which depicts that the \emph{2-competitive} nature of \emph{MFLP} with respect to \emph{STAT}. But at the same time the competitive ratio of \emph{MFM} keeps grwoing beyond the 2 factor.
\begin{table}
\centering
\begin{tabular}{ |c|c|c|c|  } 
\hline
List Size $L$ & MFLP   & MFM & MTF  \\
\hline
\emph{l=4} & \emph{2} & \emph{2} & \emph{1.6}  \\
\hline
\emph{l=6} & \emph{2.4} & \emph{2.4} & \emph{1.714}  \\
\hline
\emph{l=10} & \emph{2.5} & \emph{2.85} & \emph{1.81}  \\
\hline
\emph{l=20} & \emph{2.35} & \emph{3.33} & \emph{1.90}  \\
\hline
\emph{l=30} & \emph{2.22} & \emph{3.52} & \emph{1.93}  \\
\hline
\emph{l=40} & \emph{2.22} & \emph{3.63} & \emph{1.95}  \\
\hline
\emph{l=50} & \emph{2.17} & \emph{3.70} & \emph{1.96}  \\
\hline
\emph{l=100} & \emph{2.105} & \emph{3.84} & \emph{1.98}  \\
\hline
\emph{l=500} & \emph{2.0285} & \emph{3.968} & \emph{1.996}  \\
\hline
\emph{l=1000} & \emph{2.016} & \emph{3.99} & \emph{1.998}  \\
\hline
\emph{l=2000} & \emph{2.0090} & \emph{3.992} & \emph{1.999}  \\
\hline
\emph{l=5000} & \emph{2.0044} & \emph{3.996} & \emph{1.9996}  \\
\hline
\emph{l=10000} & \emph{2.0024} & \emph{3.998} & \emph{1.9998}  \\
\hline
\emph{l=50000} & \emph{2.00056} & \emph{3.99968} & \emph{1.99996}  \\
\hline
\emph{l=100000} & \emph{2.0003} & \emph{3.99984} & \emph{1.99998}  \\
\hline
\emph{l=1000000} & \emph{2.000036} & \emph{3.999984} & \emph{1.999998}  \\
\hline
\emph{l=100000000} & \emph{2.000001} & \emph{4} & \emph{2}  \\
\hline
 \end{tabular}
 \caption{Competitive ratio of MFLP, MFM and MTF for different size of L}
\end{table}\\
\\

{\bf Theorem:3} \emph{\bf MFLP is 2-competitive with respect to Dynamic Optimum Offline \emph{OPT} algorithm.}

\emph{\bf Proof:} Let the initial configuration of the list \emph{L} be $<x_{1}, x_{2},.., x_{p},x_{p+1},..., x_{l}>$ and the cruel request sequence be $\sigma$ of size \emph{n}. Let the cruel request sequence $\sigma$ is presented as $<x_{p}, x_{p-1},.., x_{2}, (x_{l}, x_{l-1},..,x_{p+2}, x_{p+1}, x_{1})^k >$

When the first item $x_{p}$ is accessed by \emph{MFLP}, an access  cost of \emph{p} is incurred. Then $x_{p}$ is moved to the front of the list and the item at position $(p-1)$ takes the $p^{th}$ position of the list. Again the new item at the \emph{p} position is requested and it is moved to the front of the list. This process continues for \emph{p-1} items of the request sequence till the item $x_{2}$ is accessed. For serving this subsequence i.e from item $x_{p}$ to $x_{2}$, the total cost incurred by \emph{MFLP} i.e is $p(p-1)$. 

Then the rest of subsequence i.e $<(x_{l}, x_{l-1},..,x_{p+2}, x_{p+1}, x_{1})>$ is repeated \emph{k} times in $\sigma$. The length of the subsequence is $l-p+1$ which incurs the cost \emph{l}, every time an item is accessed. So the cost incurred to serve this subsequence by \emph{MFLP} is $k[l(l-p+1)]$. So the total accessed cost incurred by \emph{MFLP} to serve the given cruel request sequence is $C_{MFLP}(L,\sigma)=$ $p(p-1)+k[l(l-p+1)]$.

The cost incurred by the dynamic optimum offline algorithm \emph{OPT} to serve the subsequence i.e from item $x_{p}$ to $x_{2}$ i.e is $p(p-1)$. \emph{OPT} incurs the cost $l(l-p+1)$ to serve the subsequence for the first time $<(x_{l}, x_{l-1},..,x_{p+2}, x_{p+1}, x_{1})>$. Then the cost incurred by \emph{OPT}
to serve the rest of subsequence which has occurred \emph{k-1} time is $(k-1)[(l-p+1)^{2}]$. Hence the total cost incurred by the \emph{OPT} to serve the cruel request sequence is $p(p-1)+l(l-p+1)+(k-1)[(l-p+1)^{2}]$.

Hence the competitive ratio of \emph{MFLP} with respect to the \emph{OPT} is $\frac{C_{MFLP}(L,\sigma)}{C_{OPT}(L,\sigma)}=$ \\$\frac{p(p-1)+k[l(l-p+1)]}{p(p-1)+l(l-p+1)+(k-1)[(l-p+1)^{2}]}$ $= \frac{p(p-1)+l(l-p+1)+(k-1)[l(l-p+1)]}{p(p-1)+l(l-p+1)+(k-1)[(l-p+1)^{2}]} \leq 2- \frac{2(l-p+1)-l}{l-p+1}$.\\

\emph{\bf Observation:} When we consider the conventional static optimum offline algorithm to perform the competitive analysis, we find it not \emph{2-competitive}. So to provide an upper bound on the cost of optimum offline algorithm, we design the dynamic optimum offline algorithm for which our \emph{MFLP} is shown to be \emph{2-competitive}.

\section*{Conclusion}
In this paper, we proposed a new deterministic online algorithm \emph{MFLP} which is the variant of \emph{MFM}. We observed that the competitive ratio of our proposed algorithm is approaching to 2 with respect to \emph{STAT} as the list size grows. However, the competitive ratio of \emph{MFM} converges to 4 for larger list size. Though the nature of both the online algorithms is similar but due to the different values of \emph{p} their behaviour differs from each other, where $\log_{2}{l} \leq p \leq m $. Our theoretical analysis shows that \emph{MFLP} has the better competitive bound as compare to the \emph{MFM}. However the experimental study can be considered as a potential future work to evaluate the performance of \emph{MFLP} on real dataset like Calgary Corpus and and Canterbury Corpus.

\end{document}